\documentclass[12pt]{iopart}

\usepackage{iopams}
\usepackage{color}
\usepackage{caption}
\usepackage[colorlinks=true,linkcolor=blue]{hyperref}
\usepackage{url}
\usepackage[dvipsnames]{xcolor}
\usepackage{adjustbox}
\usepackage{graphicx} 
\usepackage{float}
\begin{document}

\title[Optimizing beam-ion confinement in ITER by adjusting the $\Delta\varphi$ of the RMPs]{Optimizing beam-ion confinement in ITER by adjusting the toroidal phase of the 3-D magnetic fields applied for ELM control}
\author{L. Sanchis$^{1,2}$, M. Garcia-Munoz$^{2,3}$, E. Viezzer$^{2,3}$, A. Loarte$^{4}$, \\ 
L. Li$^{5,6}$, Y.Q. Liu$^{7}$, A. Snicker$^{1}$, L. Chen$^{8}$, F. Zonca$^{9}$, \\
S.D. Pinches$^{4}$, D. Zarzoso$^{10}$}

\address{{\it $^1$ Dept. of Applied Physics, Aalto University, FI-00076, Aalto, Finland}}
\address{{\it $^2$ Dept. of Atomic, Molecular and Nuclear Physics, University of Seville, Avda. Reina Mercedes, 41012 Seville, Spain}}
\address{{\it $^3$ Centro Nacional de Aceleradores CNA (Universidad de Sevilla, Junta de Andaluc\'ia, CSIC), Avda. Thomas A. Edison 7, 41092 Seville, Spain}}
\address{{\it $^4$ ITER Organization, Route de Vinon-sur-Verdon, CS 90 046, 13067 St Paul-lez-Durance Cedex, France}}
\address{{\it $^5$ College of Science, Donghua University, Shanghai 201620, China}}
\address{{\it $^6$ Member of Magnetic Confinement Fusion Research Centre, Ministry of Education, China}}
\address{{\it $^7$ General Atomics, PO Box 85608, San Diego, CA 92186-5608, United States of America}}

\address{{\it $^8$ IFTS, Zhejiang University, 310027, 310027, Hangzhou, China}}
\address{{\it $^{9}$ ENEA C.R. Frascati, CP 65-00044, Frascati, Italy}}
\address{{\it $^{10}$ Aix-Marseille Universit\'e, CNRS, PIIM, UMR 7345, Marseille, France}}

\ead{lucia.sanchis@aalto.fi}
\vspace{10pt}		

\begin{abstract} 

The confinement of Neutral Beam Injection (NBI) particles in the presence of  n=3 Resonant Magnetic Perturbations (RMPs) in 15 MA ITER DT plasmas has been studied using full orbit ASCOT simulations.  Realistic NBI distribution functions, and 3D wall and equilibria, including the plasma response to the externally applied 3D fields calculated with MARS-F, have been employed. The observed total fast-ion losses depend on the poloidal spectra of the applied n=3 RMP as well as on the absolute toroidal phase of the applied perturbation with respect to the NBI birth distribution. The absolute toroidal phase of the RMP perturbation does not affect the ELM control capabilities, which makes it a key parameter in the confinement optimization. The physics mechanisms underlying the observed fast-ion losses induced by the applied 3D fields have been studied in terms of the variation of the particle canonical angular momentum ($\delta P_{\phi}$) induced by the applied 3D fields. The presented simulations indicate that the transport is located in an Edge Resonant Transport Layer (ERTL) as observed previously in ASDEX Upgrade studies. Similarly, our results indicate that an overlapping of several linear and nonlinear resonances at the edge of the plasma might be responsible for the observed fast-ion losses. The results presented here may help to optimize the RMP configuration with respect to the NBI confinement in future ITER discharges.
 
\end{abstract}

\newpage
\section{Introduction}

In tokamak plasmas operating in H-mode \cite{hmode}, magnetohydrodynamic instabilities can lead to a large release of particles and energy to the plasma facing components. In particular, type-I edge localized modes (ELMs) \cite{elms,elms-eli}, caused by the steep gradients at the plasma edge, are one of the main concerns for ITER \cite{Loarte2003, Tani, elms-iter} and the associated power fluxes need to be controlled since they can lead to a significant reduction of plasma component lifetime and to the contamination of the plasma by the impurities produced during ELMs. Among different techniques to mitigate ELMs, the application of resonant magnetic perturbations (RMPs) has successfully mitigated or even suppressed ELM instabilities and has been adopted as the main base line technique for ITER \cite{Evans, Hender1992, Liang, Suttrop2011, Loarte2006N,Loarte2014}. However, the symmetry-breaking fields applied for ELM control can lead to a degradation of plasma energy and particle confinement, particularly of fast-ions \cite{spong-3d, Shaing2001, shaing2014,suttrop2016}. This is a concern for ITER since an excessive loss of fast particles reduces the effective Neutral Beam Injection (NBI) power available to heat the plasma and can lead to localized power fluxes \cite {Akers2016}.  
\\
\\
In most present tokamaks, RMPs are generated by two or three toroidal rows of window-frame coils where the current flowing through the coils is modulated toroidally with a given symmetry and the relative toroidal phase between the waveforms can be adjusted to vary its effect on the plasma. The relative toroidal phase between toroidal rows is thus used to modify the poloidal mode spectrum of the perturbation, which has been observed to affect the plasma stability in several experiments such as MAST \cite{kirk2010}, ASDEX Upgrade \cite{Willensdorfer2017} and DIII-D \cite{Paz-Soldan2015, Nazikian}. In the presence of these RMP fields, energetic particles are especially sensitive to this complex 3D magnetic field structure due to their relatively long mean free path and slowing down times \cite{Garcia-Munoz2013, Taina2017, Shinohara2012, VanZeeland2015,Jari2016}. However, as mentioned above, a good confinement of the energetic ions is crucial to ensure efficient heating of burning fusion plasmas since fast particles are the main source of plasma heating (NBI, alpha heating and ICRH heating) and of momentum (NBIs) into the plasma. ITER is designed on the basis of the H-mode scenario at 15 MA/5.3 T to achieve its Q = 10 goal and thus it is equipped with a set of in-vessel coils to achieve RMP fields for ELM control \cite{Loarte2006N, Loarte2014} and thus it is essential to analyze in detail the impact of RMP perturbations on the energetic particle population and to miminize it as far as possible.
\\
\\
Previous work has shown that linear and nonlinear resonant interactions between fast-ions and the RMP fields in the region near the separatrix lead to an edge resonant transport layer (ERTL) that is responsible for the observed fast-ion transport in ASDEX Upgrade \cite{ls2019}. To characterize the RMP-induced transport, the variation of the particle toroidal canonical momentum ($P_{\phi}$) caused by the broken magnetic symmetry was used as a figure of merit. The model used to calculate the particle transport was then compared and validated using the fast-ion loss measurements of the Fast-Ion Loss Detector (FILD) \cite{Garcia-Munoz2009}. In this work, the methods used to understand the fast-ion losses induced by RMPs in ASDEX Upgrade are applied to analyze ITER plasmas. By applying these models, several RMP configurations are studied in order to optimize the operational point that should ensure ELM suppression in ITER while maintaining  fast-ion NBI losses acceptable. The RMP optimization was carried out considering two main parameters, the poloidal mode spectra of the RMP perturbed fields and the absolute toroidal phase of the perturbation. While the RMP poloidal spectra can affect both the ELM control capabilities and the fast-ion confinement, the absolute toroidal phase modifies the fast-ion behaviour without altering the ELM control performance of the configuration.
\\
\\
This paper is organized as follows: Section \ref{sec:fields} presents the magnetic field structure for several RMP configurations with different poloidal mode spectra, through relative toroidal phase adjustments and thus effect on the plasma. In section \ref{sec:op}, the impact of the poloidal mode spectra of the RMP perturbation and of its absolute toroidal phase on fast-ion NBI losses is evaluated using the variation of the toroidal canonical momentum as quantifying variable. In section \ref{sec:ertl}, the wave-particle resonant interaction is presented as the transport mechanism responsible for the particle losses observed. Finally, section \ref{sec:sum} summarizes this work.


\begin{figure}[b]
    \centering
       \includegraphics[scale=0.35]{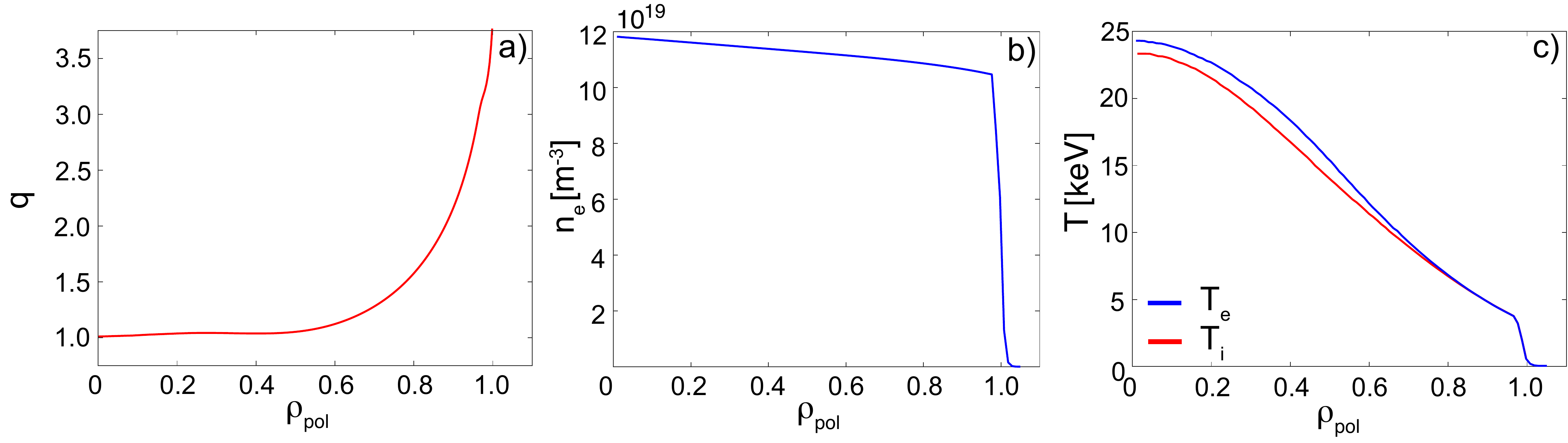}
			 \caption{\it \small ITER DT baseline scenario: a) Safety factor, b) electron density and c) electron (blue) and ion (red) temperature profiles as a function of the normalized poloidal flux coordinate $\rho_{pol}$.}
        \label{fig:profs}
\end{figure}  

\section{Perturbed magnetic fields calculated with the MARS-F code}\label{sec:fields}

In this work, the 3D plasma equilibria were generated by the combination of the axisymmetric field of ITER, and the 3D perturbation induced by the RMP coils for a range of current waveforms including the response of the plasma as evaluated with the MARS-F code.  

\begin{figure}[h]
    \centering
       \includegraphics[scale=0.65]{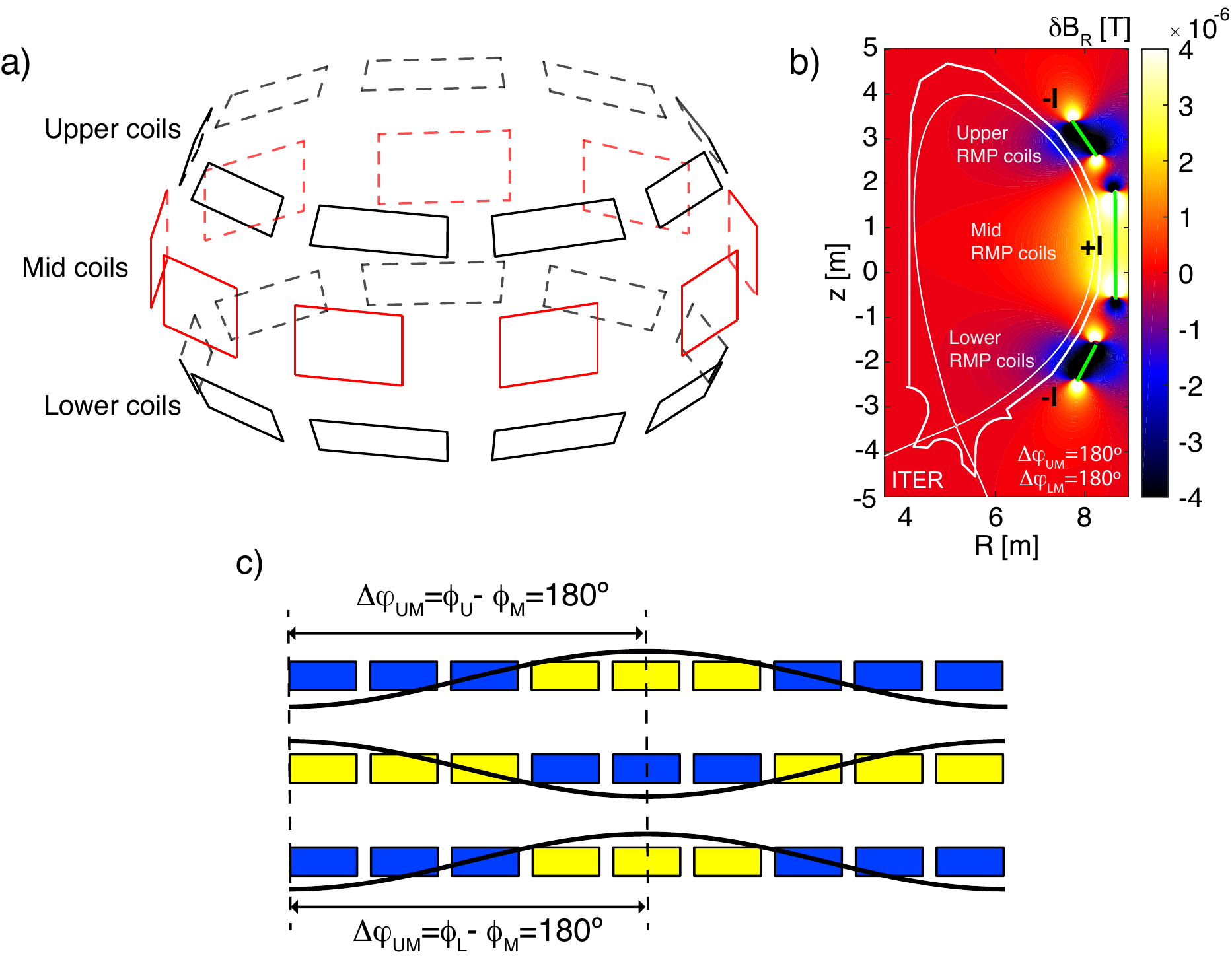}
			 \caption{\it \small a) Schematic representation of the ITER RMP coils set-up. b) Radial component of the perturbed magnetic field generated by positive (yellow) and negative (blue) 1\,A currents flowing through the RMP coils as a function of the radial and z coordinates. c) Phase shifts between the current waveforms flowing through the upper, mid and lower set of coils.}
        \label{fig:coils3d}
\end{figure} 

In ITER, the RMPs are generated by 3 toroidal rows of 9 in-vessel ELM coils, which are individually powered (Figures \ref{fig:coils3d} a) and b)). This allows changes to the absolute toroidal phase of the perturbation applied as well the relative toroidal phase of the current waveforms between the upper and equatorial and lower and equatorial rows of coils (Figure \ref{fig:coils3d} c)). For the analysis in this paper, we consider current waveforms with n = 3 symmetry and a maximum current level of 90 kAt in the coils, which is its maximum design value \cite{Loarte2014}. The poloidal mode spectra of the perturbation applied is determined by the phase shift between the upper and middle row ($\Delta\varphi_{UM}$) and the lower and middle row ($\Delta\varphi_{LM}$), while the toroidal spectrum is determined by that of the toroidal current waveform. The plasma response to the external perturbation has been generated using the MARS-F code \cite{YQ2000}, which calculates the linear response to the RMP fields considering a single fluid model and full 3D toroidal geometry. The H-mode plasma scenario considered here corresponds to that of the Q = 10 baseline scenario to be demonstrated during the Fusion Power Operation (FPO) phase of the ITER Research Plan \cite{iter-report}. Figure \ref{fig:profs} a) shows the safety factor profile corresponding to the H-mode scenario with $I_p=15\,MA$, $B_{\phi}=5.3\,T$ and $Q=10$. Figures \ref{fig:profs} b) and c) show the radial profiles of the predicted electron density and electron and ion temperature.

\begin{figure}[h]
    \centering
       \includegraphics[scale=0.6]{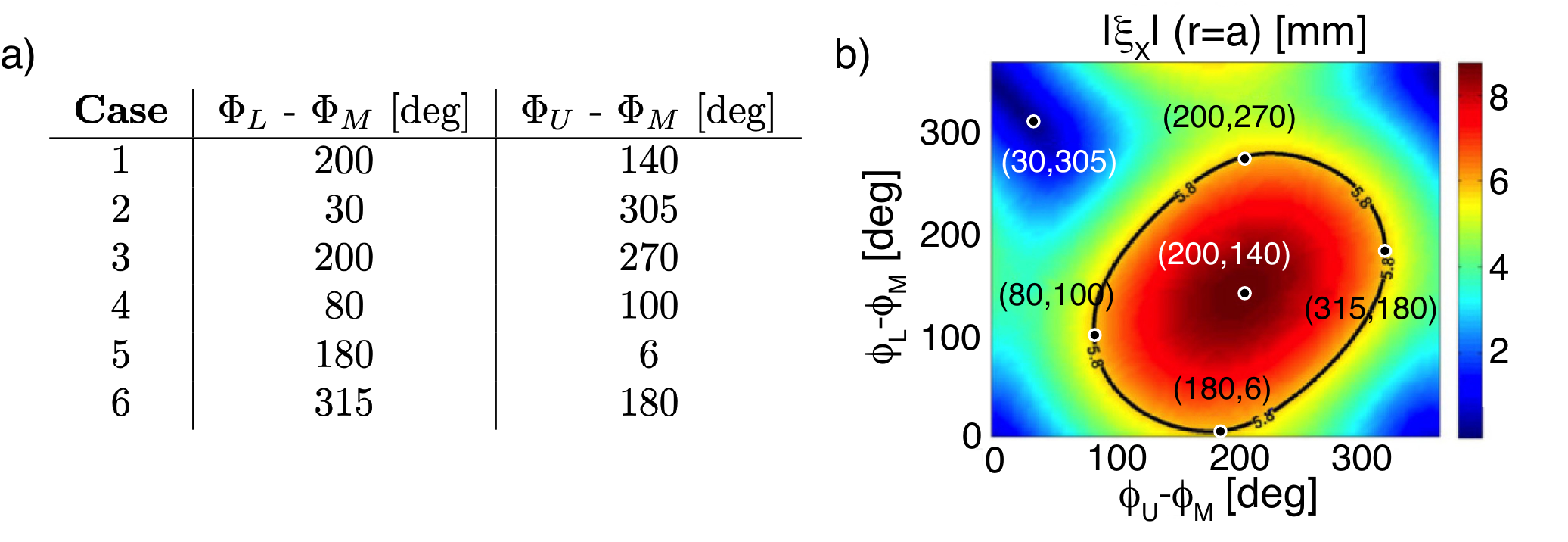}
			 \caption{\it \small a) Differential phase shift of the upper and lower set of coils with respect to the the coils in the middle row. b) The MARS-F computed plasma surface displacement near the X-point \cite{Lili}. The relative toroidal phase shifts correspond to the ones shown in a).}
        \label{fig:cases}
\end{figure}  

\begin{figure}[h]
    \centering
       \includegraphics[scale=0.5]{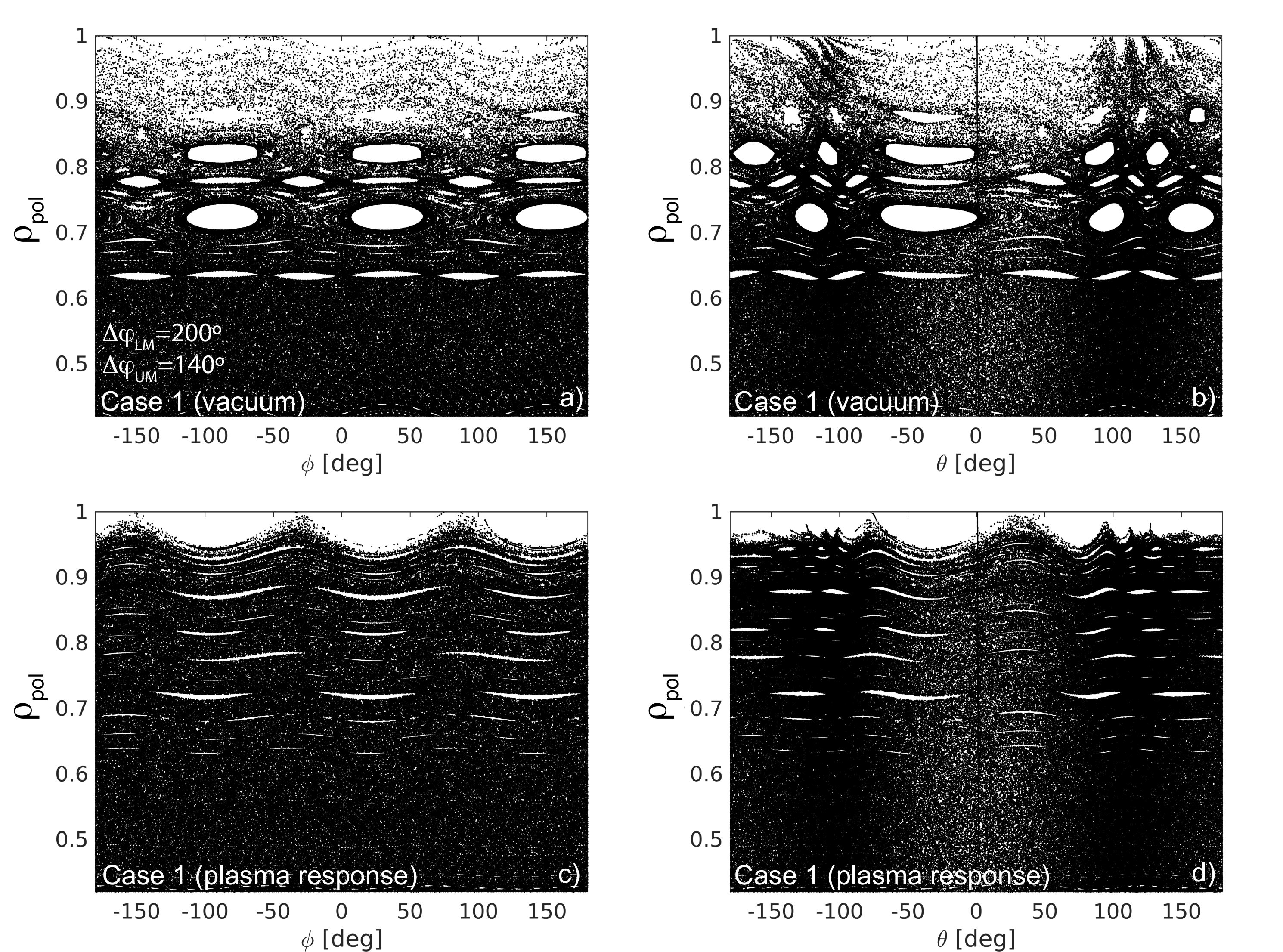}
			 \caption{\it \small Toroidal and poloidal Poincar\'e maps of the magnetic field lines for case 1 in vacuum approach (a) and b)) and including plasma response (c) and d)).}
        \label{fig:poincare_1}
\end{figure}

\begin{figure}[h]
    \centering
       \includegraphics[scale=0.5]{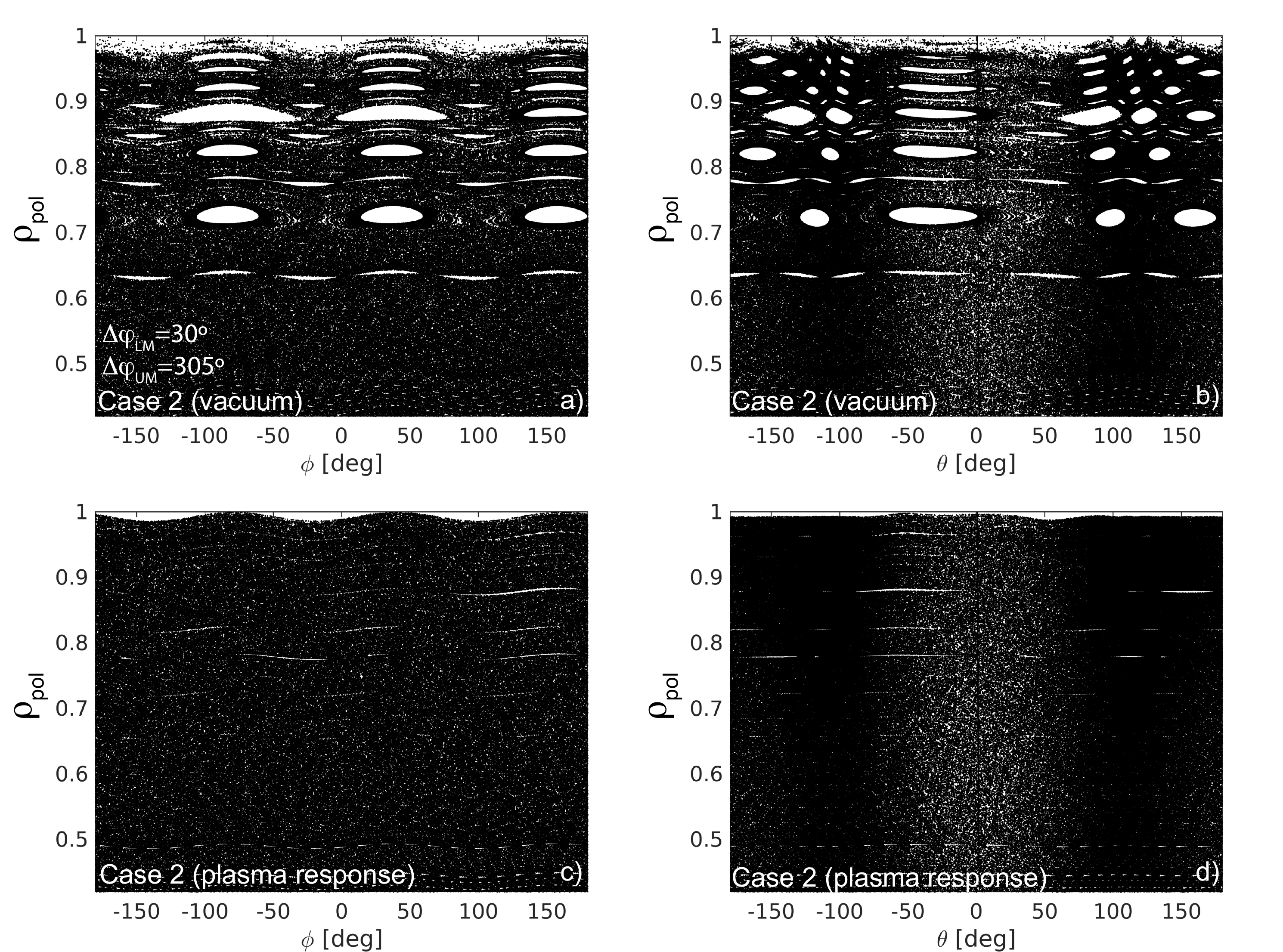}
			 \caption{\it \small Toroidal and poloidal Poincar\'e maps of the magnetic field lines for case 2 in vacuum approach (a) and b)) and including plasma response (c) and d)).}
        \label{fig:poincare_2}
\end{figure}

To vary the poloidal mode spectra of the perturbation, different relative toroidal phases have been considered. The set of RMP configurations analyzed are characterized in figure \ref{fig:cases} by the relative phase shifts between the three current waveforms and the associated plasma surface displacement at the X-point, which is a robust parameter correlated with the achievement of ELM mitigation/control by RMP fields \cite{Lili}. Case 1 corresponds to the case with maximum X-point displacement while case 2 corresponds to the minimum one for which no effect on ELM behaviour is expected from RMPs in ITER (e.g. a non-resonant case). Case 3-6 are cases for which the X-point displacement is the same but have different poloidal spectra.
\\
The poloidal and toroidal Poincar\'e maps of the magnetic field lines associated to cases 1 and 2 are shown in figures \ref{fig:poincare_1}-\ref{fig:poincare_2} in vacuum and including plasma response. Note that, in the poloidal map, the apparent lower density of magnetic field lines at $\theta=0$ is a consequence of this direction being aligned with the shortest distance between the magnetic axis and the separatrix. These two configurations show the two extreme cases in terms of RMP perturbation impact on the X-point displacement. In case 1, the resulting vacuum magnetic field shows a strongly chaotic region at the plasma edge. The toroidal Poincar\'e map in figure \ref{fig:poincare_1} a) shows $n=3$ symmetry modulated by the toroidal symmetry of the perturbation. In this case, the inclusion of plasma response decreases the effect of the perturbation reducing the width of the magnetic islands in both the poloidal and toroidal dimensions. The analysis of these Poincar\'e maps shows the chaotic regions in the magnetic field generated by the RMP perturbations and how this chaotic areas are modified by the plasma response. However, the chaotic behaviour of fast-ions in the proximity of the RMP coils is different from the magnetic field lines and so are the thresholds between the regular and chaotic regions.

\begin{figure}[h]
    \centering
       \includegraphics[scale=0.35]{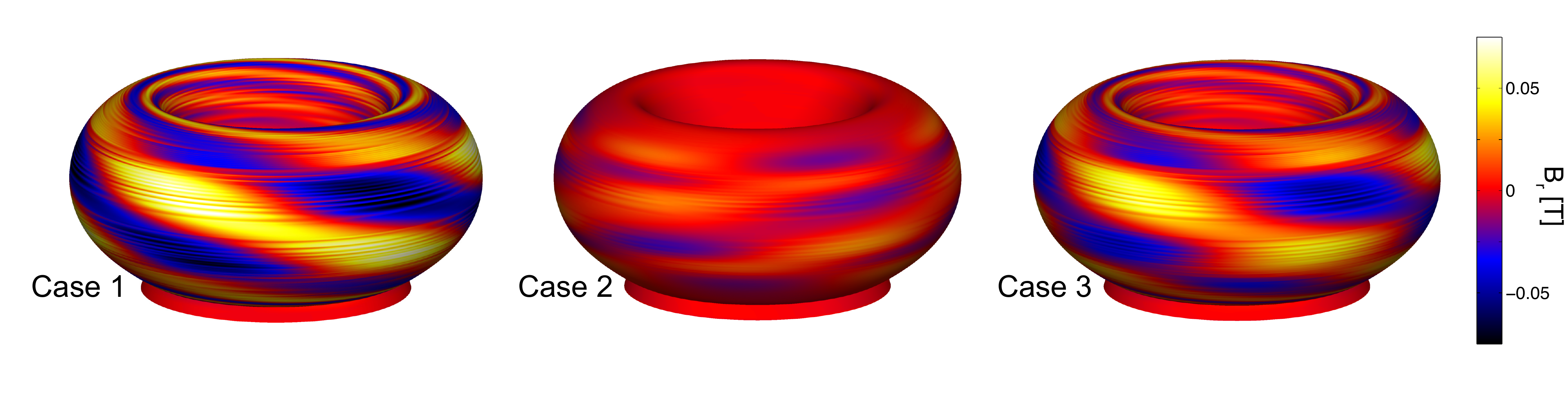}
			 \caption{\it \small Perpendicular component of the RMP perturbed magnetic field at the separatrix including the plasma response.}
        \label{fig:fields}
\end{figure}

Figure \ref{fig:poincare_2} shows similar Poincar\'e maps for case 2 (the non-resonant case) presenting a configuration with a considerable smaller chaotic region at the edge and magnetic island structures that extend to $\rho_{pol}=0.65$. The effect of the plasma response in this case leads to the suppression of the island structures and strongly reduces the chaotic behaviour at the edge. In figure \ref{fig:fields}, the component of the magnetic perturbation perpendicular to the separatrix including the plasma response is presented for cases 1-3, with 3 being a case with large X-point displacement but smaller than case 1. This figure shows that the resulting perturbed field generated using the same coil current intensity depends strongly on the relative toroidal phases of the three current waveforms in ITER. The current waveform configuration of case 1 leads to a strong perturbed field at the separatrix and has maximum X-point respond and thus effectiveness for ELM control, while case 2 has a much more reduced perturbation (reduced by a factor of 3), but also a small X-point displacement and thus no ELM control effect. The configuration associated to case 3 presents an intermediate state where ELM control should be expected.   

\section{Fast-ion loss optimization in the presence of RMPs to provide ELM control}\label{sec:op}

\begin{figure}[h]
    \centering
       \includegraphics[scale=0.65]{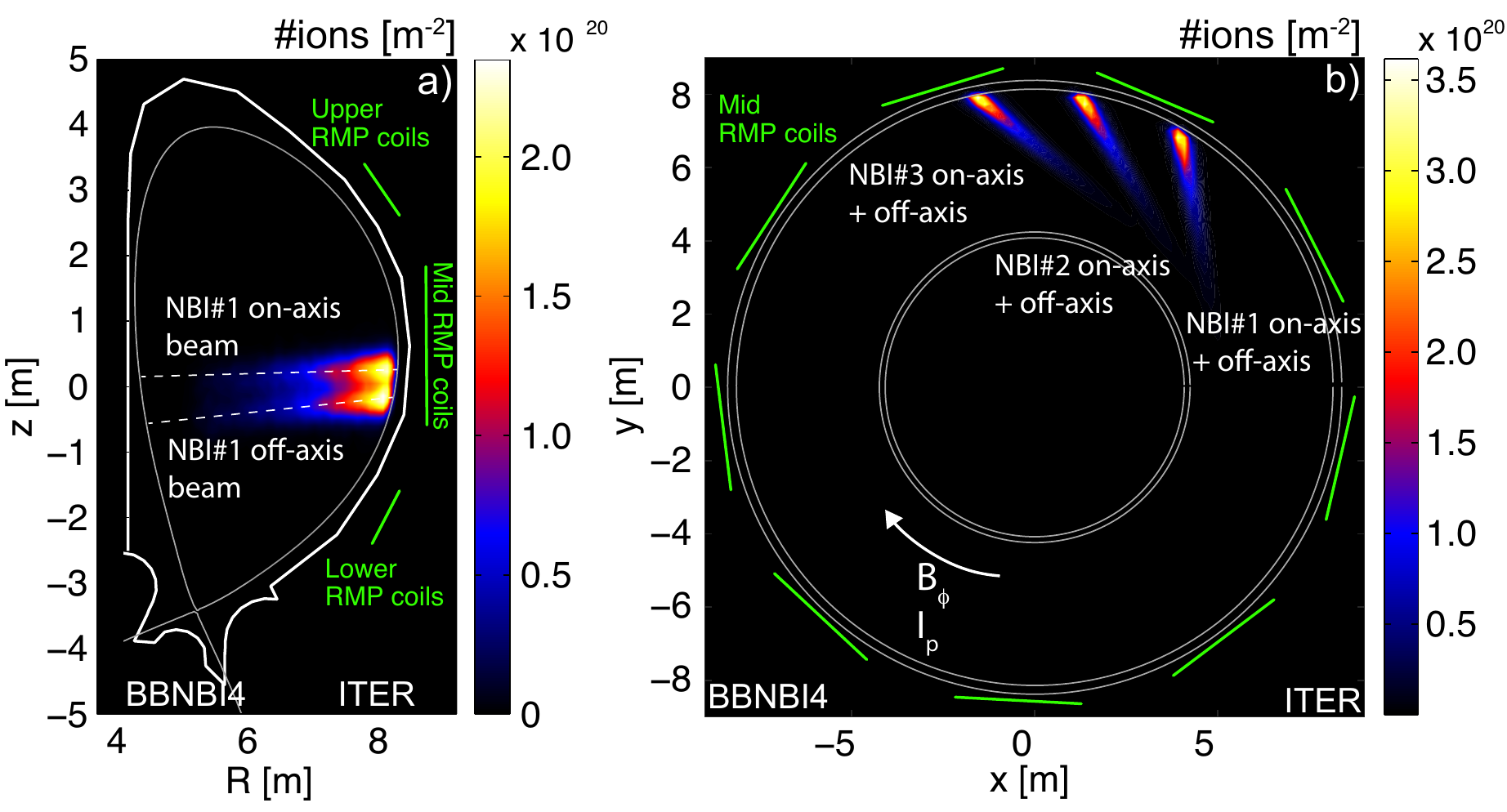}
			 \caption{\it \small 2D histogram showing the surface density of the initial NBI\#1 on and off-axis distribution projected onto the poloidal a) and toroidal b) cross sections. Green lines indicate the locations of the mid RMP coils.}
			  \label{fig:nbi}
\end{figure}

The optimal RMP configuration to reduce ion losses while ensuring ELM control is analyzed in terms of the variation of the toroidal canonical momentum ($P_{\phi}$) of the fast-ions generated by the ITER neutral beam injectors. Figure \ref{fig:nbi} shows the surface density of energetic particles from NBI injector\#1 in the poloidal and toroidal cross sections calculated with the ASCOT module BBNBI \cite{bbnbi} from the kinetic profiles shown in figure \ref{fig:profs}. The NBI system in ITER consists of up to 3 injectors (2 NBI injectors are part of the baseline with a possibility of a third one as an upgrade) located at consecutive equatorial ports, each one can be injected with on and off-axis injection geometries \cite{nbi-iter}. The NBI\#1 on-axis beam has the maximum ionization at $z=0.40$, $\phi=58^{\circ}$ with pitch $\lambda=v_{\parallel}/v=0.63$, while the off-axis beam ionization is located at $z=-0.1$, $\phi=58^{\circ}$ with pitch $\lambda=0.58$.         

\subsection{Fast-ion transport analysis}

The fast-ion transport was evaluated through the variation of $P_{\phi}$ induced by the symmetry-breaking RMP fields. In order to calculate $\delta P_{\phi}$, 40000 markers are followed in full orbit motion using the ASCOT code \cite{Hirvijoki2014}. The markers are distributed in a $200\times 200$ homogeneous grid varying the initial radial and toroidal location, which ensures a sufficient resolution in the initial coordinates to observe the $\delta P_{\phi}$ structures  according to a convergence study. The parameters of this distribution are set to the ones fixed by the injection geometry of the NBI\#1 on-axis. In this case, deuterium particles are started at $z=0.40\,m$, with energy $E=1\,MeV$ and pitch angle $\lambda=0.63$, which lead to a maximum of trapped ions at $R=8.15\,m$ with the trapped/passing boundary located at $R=7.9\,m$.
\\
\indent The calculation of the variation of $P_{\phi}(i)$ is done by averaging the evolution of $P_{\phi}(i)$ along the ion trajectory for approximately 10 poloidal turns or until the ion hits the wall using the expression:

\begin{equation}
 \langle \delta P_{\phi}\rangle=\sum_{i=1}^N\frac{P_{\phi}(i)-P_{\phi}(0)}{N}
\label{eq:pphi}
\end{equation} 

\begin{figure}[h]
    \centering
       \includegraphics[scale=0.35]{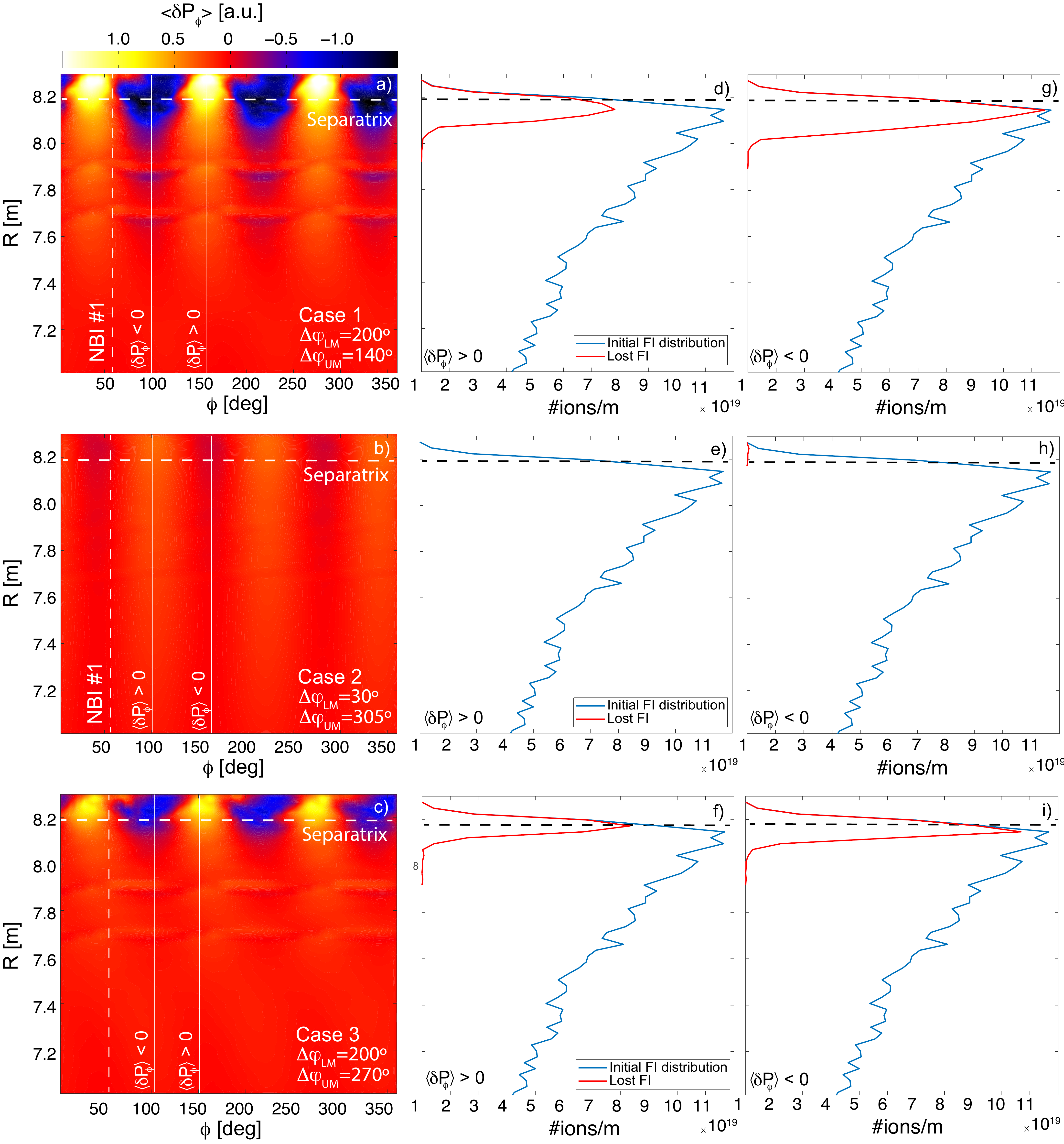}
			 \caption{\it \small a)-c) $\langle\delta P_{\phi} \rangle$ as a function of the particle toroidal angle and major radius for cases 1-3. d)-i) Radial histogram of the initial radial distribution of the beam NBI\#1 on axis (blue) and lost fast-ions (red) for the initial toroidal angle leading to a minimum (d)-f))  and maximum (g)-i)) fast-ion losses.}
        \label{fig:dp_phi_1}
\end{figure}

Here, $P_{\phi}(i)$ is the value of $P_{\phi}$ at each time point considered to evaluate $\langle \delta P_{\phi}\rangle$ and $N$ is the total number time points along the trajectory. $P_{\phi}$ is a constant of motion in 2D fields, but varies when a 3D perturbation is applied. This variation can be related to a radial transport of the particle where the direction of the drift depends on the sign of $\langle \delta P_{\phi}\rangle$. By considering the sign of the particle toroidal velocity $v_{\phi}$ and poloidal flux $\psi$ positive in the direction of the plasma current (as shown in figure \ref{fig:nbi} b)), a positive variation of $\langle\delta P_{\phi}\rangle$ leads to an inwards transport while a negative variation implies an outwards transport. Most of the fast NBI ions that are lost come from the edge as shown in figure \ref{fig:dp_phi_1}. In this region, the typical ion trajectory in the presence of RMPs is lost within 4-8 poloidal turns. This ensures that the following time used to calculate $\langle \delta P_{\phi}\rangle$, which covers approximately 10 poloidal turns, represents a good approximation to the variation of $P_{\phi}$ caused by the RMP fields.

Using this variation, the global toroidal phase of the perturbation was scanned (note that maximum deposition of the NBI injector\# 1 is located at a toroidal angle of 58 degrees)  and the maximum radial transport evaluated for all 6 cases. In figures \ref{fig:dp_phi_1} a)-c), $\langle\delta P_{\phi}\rangle$ is presented as a function of the particle initial radial position and starting toroidal angle at fixed initial $z=0.40\,m$, $E=1\,MeV$ and $\lambda=0.63$ for three representative cases. Case 1 associated to the largest X-point displacement also leads to the maximum variation of $\langle\delta P_{\phi}\rangle$. For case 2, with minimum X-point displacement, there is a smaller variation of $\langle\delta P_{\phi}\rangle$. Case 3 has a sizeable X-point displacement but smaller than case 1 and presents an intermediate variation of $\langle\delta P_{\phi}\rangle$ between cases 1 and 2. The transport is localized at the edge near the separatrix and the $n=3$ symmetry pattern is given by the toroidal current waveform of the applied perturbation. The dashed vertical white line indicates the location of the NBI\#1 injection and the two solid lines show the toroidal phase corresponding to the maximum inwards/outwards transport. By changing the absolute toroidal phase of the RMP perturbation applied to align the maximum variation of $\langle\delta P_{\phi}\rangle$ with the NBI\#1 injection, fast-ion losses can be analyzed for the two extreme situations leading to a radial transport in opposite directions. The absolute toroidal phase of the perturbation can be modified by performing a solid rotation of the three current waveforms flowing through the three sets of coils in the toroidal direction. 

\begin{figure}[h]
    \centering
       \includegraphics[scale=0.37]{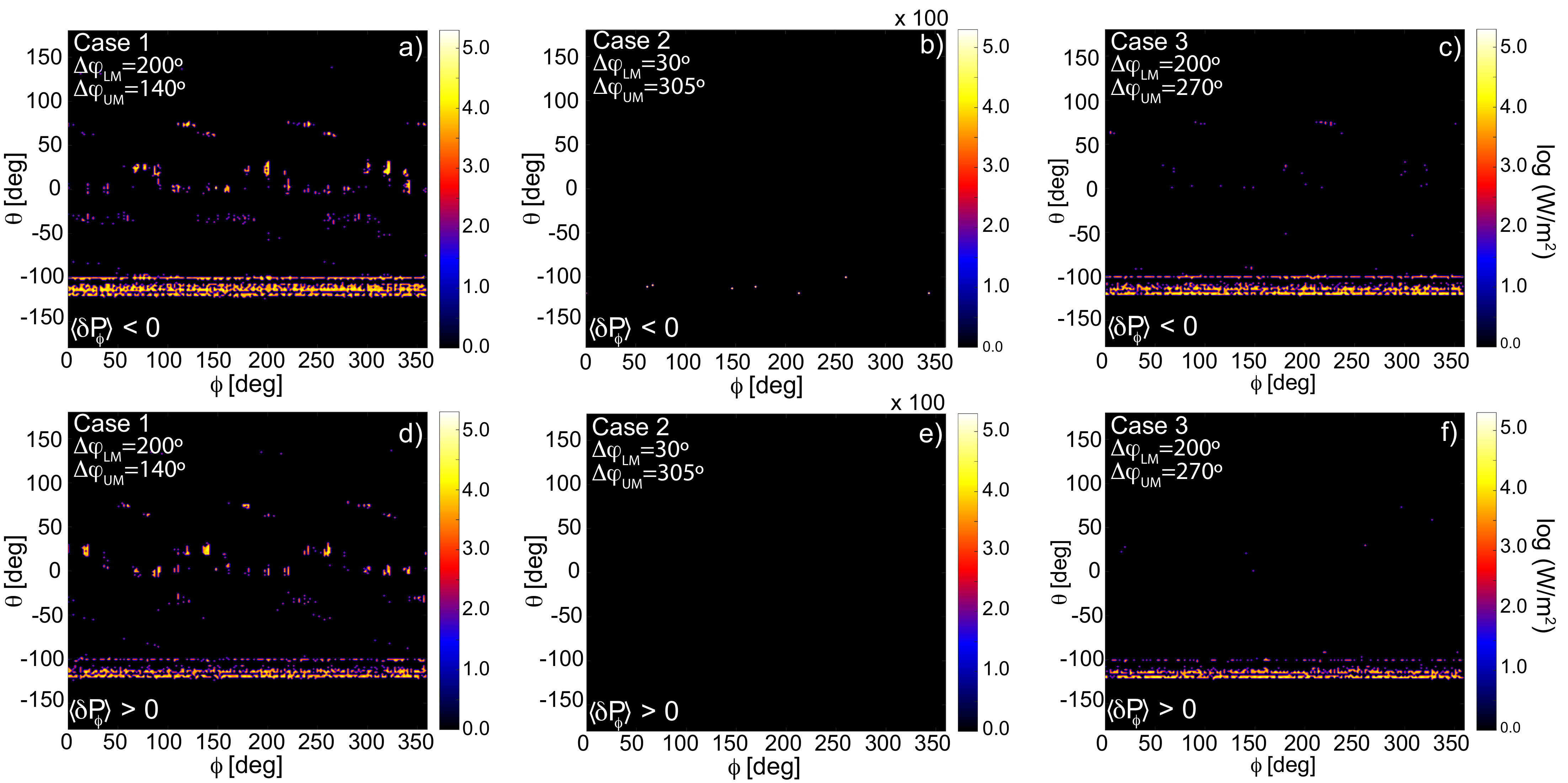}
			 \caption{\it \small Wall heat loads for cases 1-3 due to the fast-ion losses corresponding to NBI\#1 on-axis. Figures a)-c) correspond to the toroidal phase where outward transport is maximized ($\langle\delta P_{\phi} \rangle<0$) and d)-e) correspond to the toroidal phase where inward transport is maximized ($\langle\delta P_{\phi} \rangle>0$). Note that for case 2, the signal has been multiplied by 100 as the values of the heat loads are very small.}
        \label{fig:heat_1}
\end{figure}   

\begin{figure}[h]
    \centering
      \includegraphics[scale=0.37]{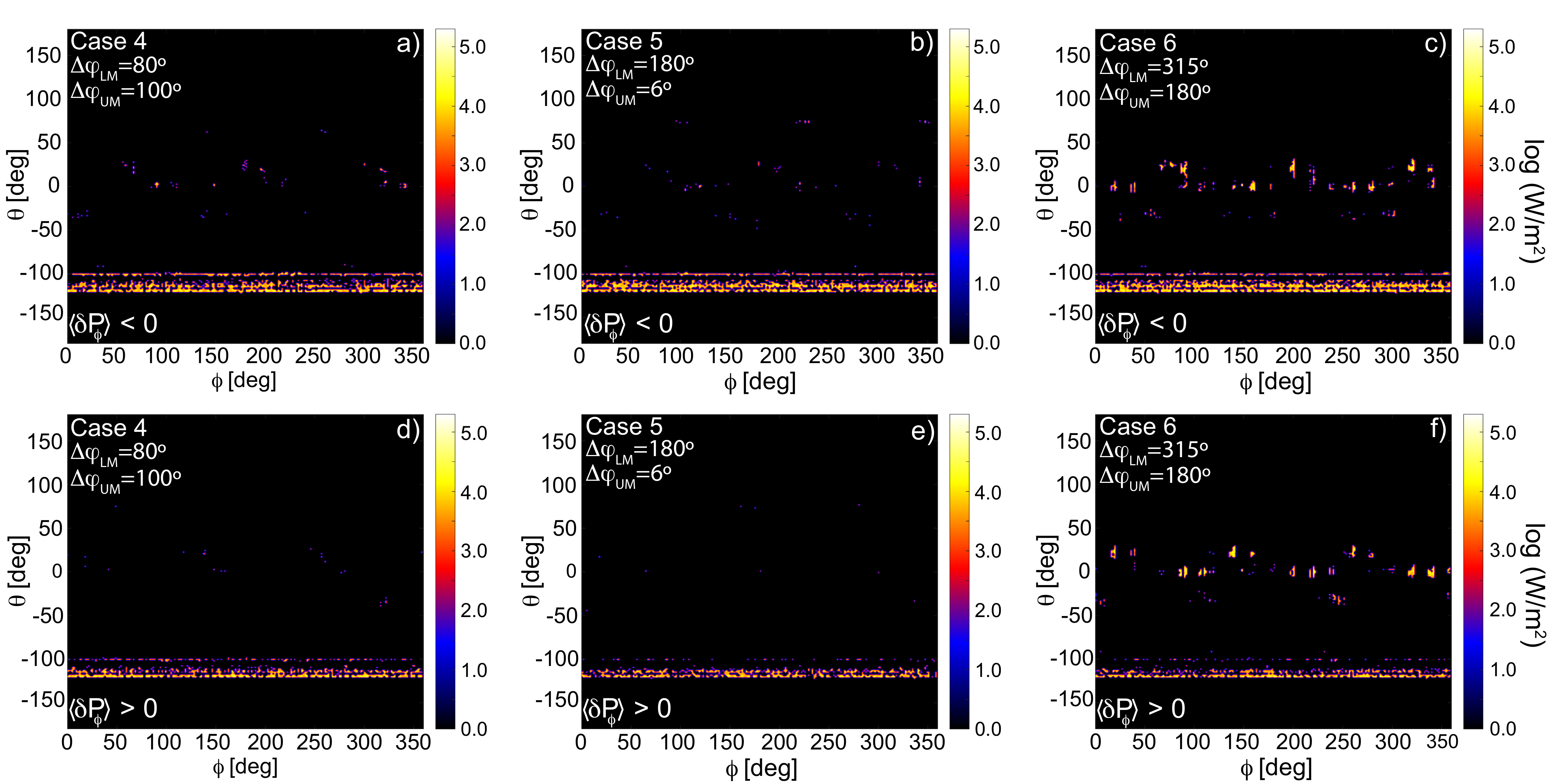}
\caption{\it \small Wall heat loads for cases 4-6 due to the fast-ion losses corresponding to NBI\#1 on axis. Figures a)-c) correspond to the toroidal phase where outward transport is maximized ($\langle\delta P_{\phi} \rangle<0$) and d)-e) correspond to the toroidal phase where inward transport is maximized ($\langle\delta P_{\phi} \rangle>0$).}
        \label{fig:heat_4}
\end{figure} 

Using a realistic NBI\#1 initial distribution consisting of 30000 markers, the fast-ion losses have been calculated for each toroidal phase. Each marker was followed in full orbit motion using a modified leap-frog integration method that ensures the energy conservation in the absence of collisions and radial electric field \cite{anttileap}, which were not considered in these simulations. The markers were followed for 20 ms with a time step $\Delta t_{step}=2\cdot 10^{-9}\,s$. Figures \ref{fig:dp_phi_1} d)-i) show the radial histogram of the initial beam deposition (blue) and the initial distribution of the lost ions (red) for the absolute toroidal phase leading to a maximum outwards and inwards particle transport. All the different current waveform configurations were simulated for the same particle initial distribution (blue) so that the only effect on the fast-ion losses is the RMP field. The maximum fast-ion losses occur for the absolute toroidal phase of the perturbation which aligns the NBI injection with the region where $\langle\delta P_{\phi} \rangle <0$, while minimum losses correspond to the $\langle\delta P_{\phi} \rangle >0$ area. These radial histograms also show that the lost ions are born at the edge region in all cases, where the variation of $\langle\delta P_{\phi} \rangle$ is most intense. Case 2 leads to a clear minimum of fast-ion losses for both toroidal phases as was expected from the small variation of $\langle\delta P_{\phi} \rangle$ consistent with the small X-point displacement.

The heat load distributions of the lost fast-ions on a 3D mesh including the main elements of the wall are shown in figures \ref{fig:heat_1} and \ref{fig:heat_4}. From these results it can be observed that the different cases not only modulate the total fast-ion losses, but also change the distribution of the loads between main wall and divertor.

\begin{table}[h]
\begin{center}
\scalebox{0.75}{
    \begin{tabular}{c|c|c|c}
      \textbf{Case} & \textbf{Maximum fast-ion losses} & \textbf{Minimum fast-ion losses} &\textbf{Relative decrease}\\
      & ($\langle\delta P_{\phi} \rangle$ $<$ $0$) & ($\langle\delta P_{\phi} \rangle$ $>$ $0$) &  \textbf{of fast-ion losses}\\
      \hline
      1 & 13.3 (\%) & 7.7 (\%) & 39\% \\
      2 & 0.03 (\%) & 0.00 (\%) &  -\\
      3 & 8.1 (\%)& 5.8 (\%) & 28\% \\
      4 & 8.4 (\%) & 6.1 (\%) & 27\% \\ 
      5 & 9.2 (\%) & 4.2 (\%) & 54\% \\ 
      6 & 12.6 (\%)& 7.7(\%) & 39\% \\ 
    \end{tabular}}
        \caption{Fraction of fast-ion losses for cases 1-6 for the two absolute toroidal phases of the RMP perturbation  that maximizes and minimizes the fast-ion outward transport and the ion loss relative decrease.}
            \label{tab:loss}
 \end{center}
\end{table}

\begin{figure}[h]
    \centering
      \includegraphics[scale=0.5]{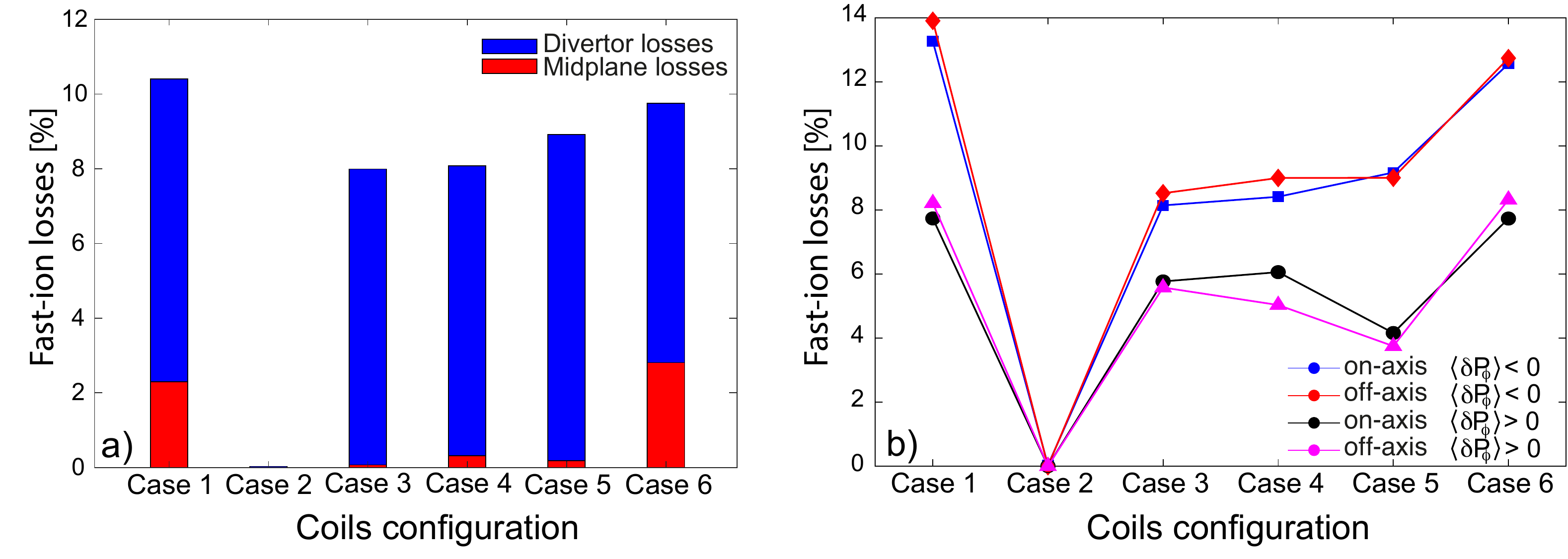}
\caption{\it \small a) Distribution of the fast-ion losses between mid plane and divertor for the on-axis case with the toroidal phase leading to the maximum negative variation of $\langle\delta P_{\phi} \rangle$. b) Fraction of lost fast-ions as a function of the perturbation poloidal mode spectra including the on and off-axis beam injections for the toroidal phase leading to the maximum variation of $\langle\delta P_{\phi} \rangle$. }
        \label{fig:totlos}
\end{figure}

The fraction of fast-ion losses integrated over the whole 3D wall is presented in table \ref{tab:loss} for NBI\#1 on axis for the 6 cases. The second column presents the particle losses considering the absolute  toroidal phase of the RMP perturbation that maximizes the outward transport ($\langle\delta P_{\phi} \rangle$ $<$ $0$). In the third column, the fast-ion losses are calculated using the toroidal phase that maximizes the inward transport and the forth column shows the relative ion loss decrease achieved. This comparison shows that the difference in the fast-ion losses depends on the absolute toroidal phase of the perturbation and can vary significantly up to almost factor of 2. This is a key input for the optimization of the RMP current waveform for ELM control, since the absolute phase does not have any impact on ELM control and thus it is an important parameter for optimization. Another observation is that RMP configurations with the same capabilities for ELM control, such as cases 3-6 which have the same X-point displacement (Figure \ref{fig:cases} b)), can have very different fast-ion losses. In some cases, these differences in the fast-ion losses can go up to almost a factor of 3, depending on the absolute RMP toroidal phase and its poloidal spectrum. Figure \ref{fig:totlos} a) shows the fraction of fast-ion losses separating between the particles hitting the midplane region and the divertor for each configuration. By comparing cases 5 and 6, which have a similar fraction of particle losses for the de-optimized absolute toroidal phase, it is clear that the poloidal mode spectra of the RMP can modify significantly the distribution of the heat loads on the main wall and divertor without changing the total losses. This is again  key input for the optimization of the RMP current waveform for ELM control, since the power handling capability of divertor (tungsten) and main wall plasma facing components is very different (beryllium) in ITER; this capability also depends on the exact location of the lost fast-ion impact within the divertor and first wall since not all divertor components (targets, baffles and dome) and first wall components have the same power handing capability. 
\\
\indent The same analysis was made for the NBI\#1 off-axis injection. Using the same absolute toroidal phases for the RMPs as for the on-axis beam, the fast-ion losses were calculated for the 6 cases. Figure \ref{fig:totlos} b) summarizes the fraction of total fast-ion losses for the two possible beam injections at each current waveform configuration. Although there are slight differences between on and off-axis, both injections show the same trend with the RMP configuration.  

\section{Identification of resonances responsible for fast-ion transport}\label{sec:ertl}

\begin{figure}[h]
    \centering
       \includegraphics[scale=0.5]{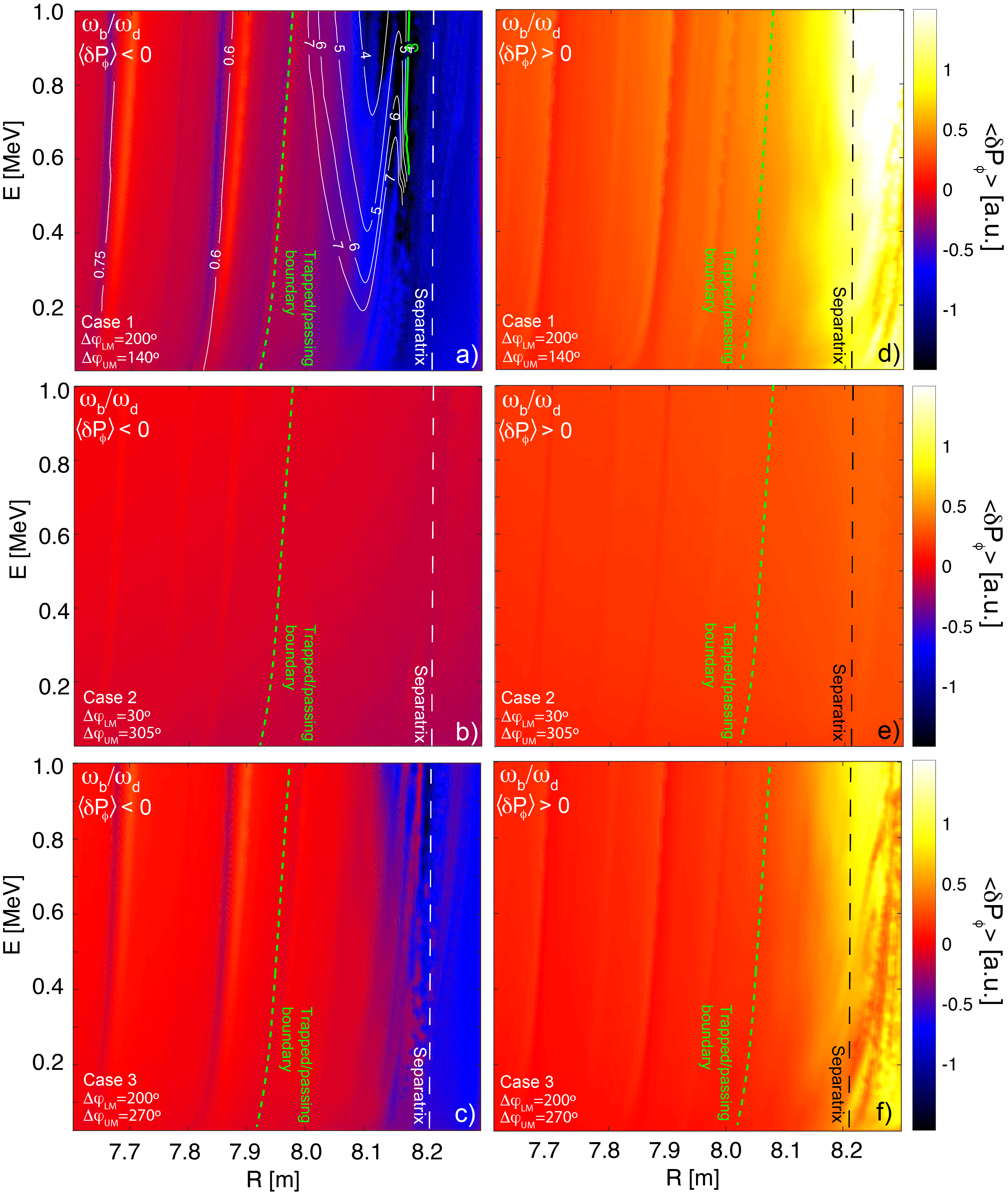}
			 \caption{\it \small $\langle\delta P_{\phi} \rangle$ as a function of the particle major radius and energy for cases 1-3 at $z=0.40\,m$ and pitch angle $\lambda=0.63$. a)-c) corresponds to the initial toroidal phase leading to maximum outward transport ($\langle\delta P_{\phi}\rangle>0$) and d)-e) to the maximum inward transport ($\langle\delta P_{\phi}\rangle<0$).}
        \label{fig:dp_er_1}
\end{figure}

\begin{figure}[h]
    \centering
       \includegraphics[scale=0.5]{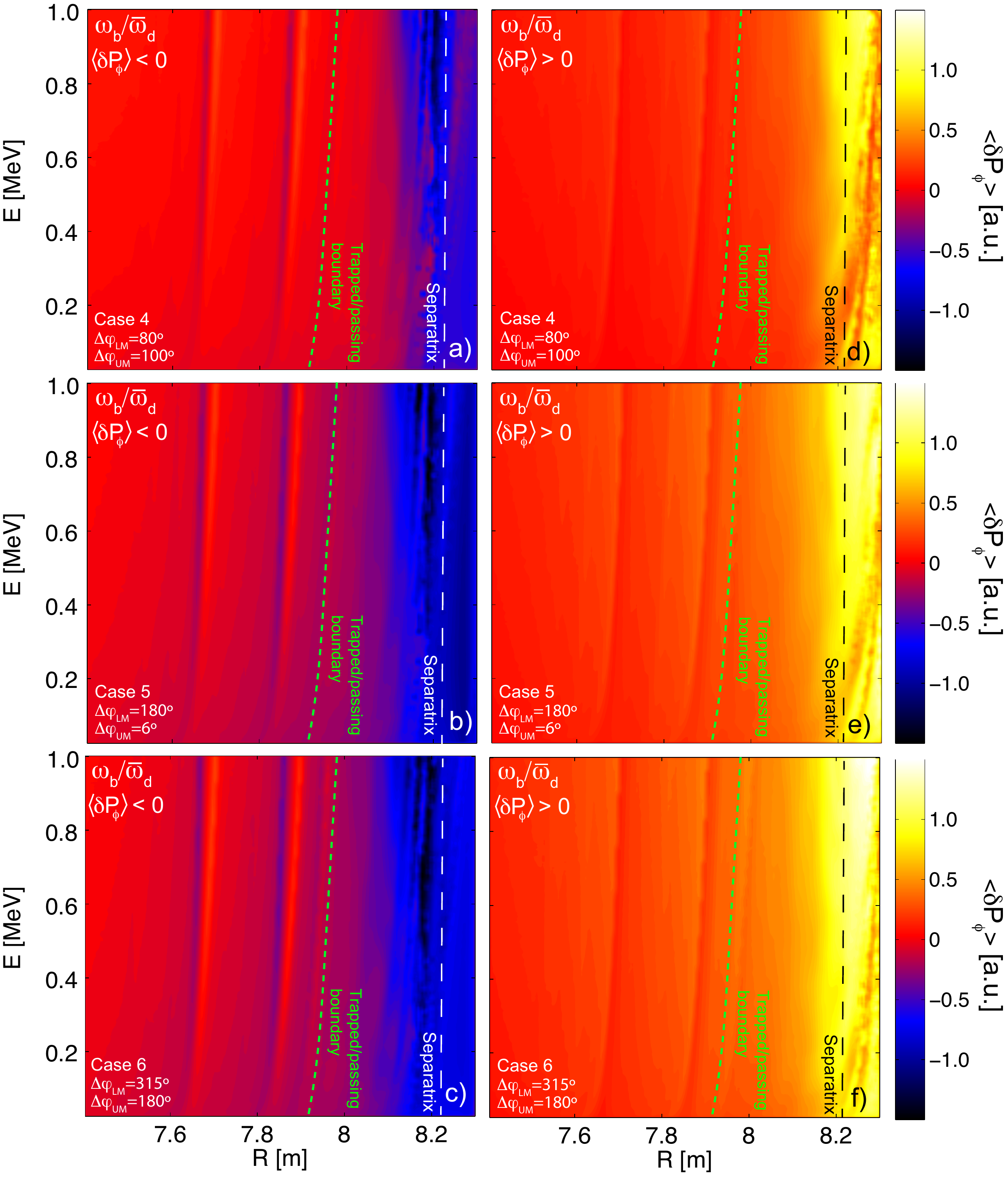}
			 \caption{\it \small $\langle\delta P_{\phi} \rangle$ as a function of the particle major radius and energy for cases 4-6  at $z=0.40\,m$ and pitch angle $\lambda=0.63$. a)-c) corresponds to the initial toroidal phase leading to maximum outward transport ($\langle\delta P_{\phi}\rangle>0$) and d)-e) to the maximum inward transport ($\langle\delta P_{\phi}\rangle<0$).}
        \label{fig:dp_er_4}
\end{figure}

In section \ref{sec:op} it was shown that the fast-ion losses are caused by RMP induced transport at the plasma edge. To further analyze this, the variation of $P_{\phi}$ was calculated as a function of the energy and the particle major radius at $z=0.40\,m$ and pitch angle $\lambda=0.63$, where the upper limit of the energy range corresponds to the NBI injection energy. Figures \ref{fig:dp_er_1}-\ref{fig:dp_er_4} show this variation for the 6 cases considering the initial toroidal angle of the NBI aligned with a negative variation of $P_{\phi}$ (a)-c)) and with a positive variation of $P_{\phi}$ (d)-f)) with respect to the initial value $P_{\phi}(t=0)$.     

Figures \ref{fig:dp_er_1} a) and d) show that case 1 leading to a maximum in the fast-ion losses corresponds to the most intense $\langle\delta P_{\phi} \rangle$ with respect to the other cases, while figures \ref{fig:dp_er_1} b) and e), corresponding to case 2, show almost no transport for all the analyzed energy range. The $\langle\delta P_{\phi} \rangle$ structures in these plots have a small dependency with the energy and its patterns can be explained by calculating the particle bounce ($\omega_b$) and precession ($\bar{\omega}_{d}$) frequencies. In figure \ref{fig:dp_er_1} a) the frequency ratio ($\omega_b/\bar{\omega}_{d}$) is indicated with white contours matching the $\langle\delta P_{\phi} \rangle$ structures. This matching indicates that the variation of $P_{\phi}$ and consequent radial transport is caused by a resonant interaction between the fast-ions and the perturbed magnetic field. Several observations can be made from this figure. First, there are two different regions that can be distinguished by the behaviour of the $\omega_b/\bar{\omega}_{d}$ structures. The region at $R<7.9\,m$ presenting vertical resonances corresponds to passing orbits, while $R > 7.9\,m$ is populated by trapped particles. Also, the overlapping of particle frequencies at the edge leads to a maximum variation of $P_{\phi}$, causing the radial transport to be localized near the separatrix. Most transport occurs for trapped ions at the edge, but there are resonances for passing orbits as well at $R=7.6-7.9\,m$. However, the effect of these resonances is not relevant as the $\langle\delta P_{\phi} \rangle$ intensity is lower since the intensity of the field created by the RMPs at this location decreases as the distance to the ELM control coils increases. Moreover, the ionization profiles of the NBI peak at the plasma edge, so the ion population affected by these interactions is smaller. From the comparison of figure \ref{fig:dp_er_1} a) and \ref{fig:dp_er_1} d), both made for case 1, it is clear that the initial relative toroidal phase has a strong impact on the ion transport. The intensity and the sign of $\langle\delta P_{\phi} \rangle$ as well as the radial location
of the resonances involved in the transport are modified.
 
\begin{figure}[h]
    \centering
       \includegraphics[scale=0.54]{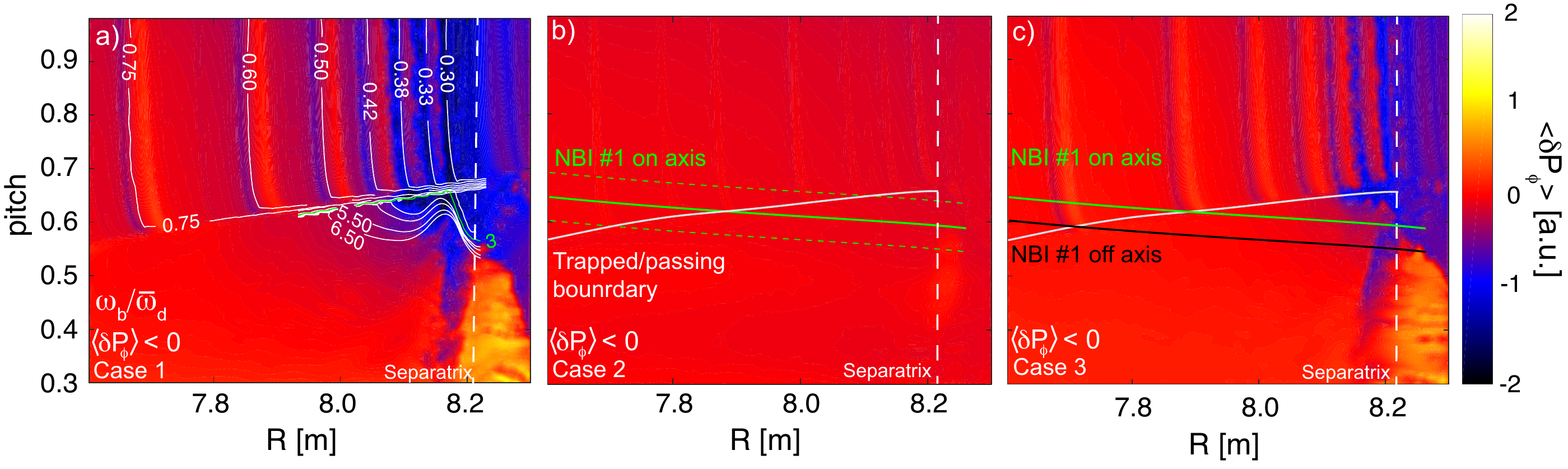}
			 \caption{\it \small $\langle\delta P_{\phi} \rangle$ as a function of the particle major radius and pitch angle for cases 1-3 at $z=0.40\,m$, toroidal angle $\phi=58^{\circ}$ and energy $E=1\,MeV$. White contours in a) indicate the constant ratio of $\omega_b/\bar{\omega}_{d}$ and the white solid line in b) and c) shows the position of the boundary between passing and trapped orbits. The NBI\#1 on-axis birth distribution is indicated with a solid green line and and the FWHM with a dashed green line. The solid black line in c) indicates the NBI\#1 off-axis birth distribution.}

        \label{fig:pitchR}
\end{figure}  

The dependence of the resonant transport with the particle initial pitch angle is shown in figure \ref{fig:pitchR} for cases 1-3. In figure \ref{fig:pitchR} a), the white contours indicate the particle resonances, matching the structure of $\langle\delta P_{\phi} \rangle$. At the trapped/passing boundary, the overlapping of resonances increase the particle transport at the plasma edge ($R> 8.0\,m$). The $\langle\delta P_{\phi} \rangle$ patterns show that the resonant transport depends strongly on the particle pitch angle for trapped particles ($\lambda<0.66$ at $R=8.2\,m$) while passing particles ($\lambda>0.66$ at $R=8.2\,m$) are nearly independent. Figure \ref{fig:pitchR} b) shows the pitch distribution of the  NBI\#1 on-axis deposition with respect to the trapped/passing boundary and the $\langle\delta P_{\phi} \rangle$ structures including the beam FWHM. In figure \ref{fig:pitchR} c), the NBI\#1 birth distribution is indicated for the on and off-axis geometries. The information of the NBI pitch distribution with respect to the perturbation can be used to determine the optimal NBI geometry for a given RMP coil configuration to reduce the particle transport at the plasma edge.         

\noindent To identify the resonances responsible for the fast-ion losses the particle resonances can be expressed as \cite{Zonca2015,chen2019,ls2019}:

\begin{equation}
\frac{\omega_{b}}{\bar{\omega}_{d}}=\frac{n(l+1)}{p(l+1)+p'}
\end{equation}

where $n$ is the toroidal mode number, $l$ is the nonlinear harmonic, $p$ is the primary bounce harmonic, and $p'$ is the nonlinear bounce harmonic.  
\\
\\
The strongest variation of $P_{\phi}$ is located in the trapped region and is aligned with $\omega_b/\bar{\omega}_{d}=3$, which corresponds to the linear resonance associated to $n=3$, given by the perturbation geometry and $p=1$. This resonance is observed in all cases shown in figures \ref{fig:dp_er_1} and \ref{fig:dp_er_4} for the $\langle\delta P_{\phi} \rangle<0$ configuration, excluding case 2. The other resonant structures can be explained by an overlapping of nonlinear resonances in the range $\omega_b/\bar{\omega}_{d}=4-7$, which shows a good agreement with the 
$\langle\delta P_{\phi}\rangle$ patterns (figure \ref{fig:dp_er_1} a)).


\section{Summary and Conclusions}\label{sec:sum}

Using the variation of the toroidal canonical momentum, the fast-ion transport in the presence of RMP fields was analyzed for the ITER Q = 10 H-mode baseline scenario with 3-D fields applied for ELM control by the in-vessel ELM control coils. This approach has been previously applied and validated in ASDEX Upgrade against experimental data, revealing the existence of an ERTL responsible for the radial particle transport. In this work, the same method was used to identify the fast-ion transport and to optimize the absolute toroidal phase of the RMP perturbation as well as its poloidal mode spectra for ITER.    
\\
The analysis of the ion transport in terms of $\langle\delta P_{\phi} \rangle>0$ for ITER showed an ERTL similar to the one previously observed in AUG. The results presented here indicate that, in this edge layer, a resonant interaction between the energetic particles and the perturbed magnetic fields may lead to the fast-ion transport observed in these ASCOT full orbit simulations.
\\
The ERTL is located within 10\,cm around the separatrix and has a strong dependency on the RMP perturbation applied.
\\
First, the absolute toroidal phase of the applied RMP perturbation has been studied. This was done by aligning the absolute phase of the perturbation with the toroidal angle at which the ITER NBI injectors are located so as to provide a maximum variation of $P_{\phi}$. The alignment with the negative variation of $P_{\phi}$ corresponded to the maximum of the outward transport while the positive variation of $P_{\phi}$ maximized the inward transport. By comparing the total fast-ion losses in each case, it has been found that the optimization of the toroidal angle can decrease the ion losses down to less than 8\% for the largest X-point displacement configuration studied, which corresponds to more than 50\% decrease compared to the de-optimized phase. The analysis to make this first optimization is fast in terms of CPU-time and the results can be directly applied by changing the absolute  toroidal angle of the RMP perturbation since the NBI location is fixed. This optimization has no impact on the capabilities of the RMP perturbation applied to provide ELM control.
\\
In the second step of this analysis, the poloidal mode spectra was optimized using the variation of $P_{\phi}$ to scan  different configurations that have the same potential for ELM control, i.e. the same X-point displacement. The analysis showed that fast-ion losses are not necessarily correlated with X-point displacement and  the associated losses for cases 3-6 showed a relative variation of up to $\pm$40\%, once the absolute RMP phase is optimized to minimize losses, potentially down to 4.2\% for case 5, which is a very low value for the ELM suppressed H-modes required in ITER. 
\\
Finally, by comparing $\langle\delta P_{\phi} \rangle$ with the particle resonances ($\omega_b/\bar{\omega}_{d}$), it was shown that the observed particle transport in our simulations could be explained by an overlapping of multiple linear and nonlinear resonances close to the separatrix. Both linear and nonlinear resonances are involved in the ERTL, however, the linear resonance $\omega_{b}/\bar{\omega}_d=3$ ($n=3$, $p=1$, $l=0$) is dominant. This suggests that the toroidal symmetry of the RMP perturbation also entails a significant impact on the intensity and location of the fast-ion transport and thus other RMP configurations that are planned to be used in ITER for ELM control (n = 4 and, potentially, n = 2) should also be analyzed as well as other plasma scenarios. From the results in our studies, the effects on alpha particles are expected to be much smaller given that their source is in the plasma central area.
\\
The procedure developed here to select the optimal RMP configuration can be applied to any plasma equilibrium and profiles and particle species to define the best ELM control scheme to target ELM suppression while avoiding the degradation of the fast-ion confinement. Our study also highlights the need to include appropriate diagnostics for fast particle losses in ITER since they are likely to be required for the optimization of the operational scenarios and RMP characteristics to ensure ELM suppression, good fast-ion confinement and acceptable fast-ion power fluxes to the main wall and divertor in ITER.      

\section*{Acknowledgments}

This work has received funding from the Spanish Ministry of Science, Innovation and Universities (grant BES-2013-065501). The simulations were partly performed on the MARCONI supercomputer (CINECA) under project reference FUA33\_ASCOT-US. This work has been partially carried out within the framework of the EUROfusion Consortium and has partially received funding from the Euratom research and training programme 2014–2018 and 2019–2020 under grant agreement No. 633053. The views and opinions expressed herein do not necessarily reflect those of the European Commission. The support from the European Research Council (ERC) under the European Union’s Horizon 2020 research and innovation programme (grant agreement No. 805162) is gratefully acknowledged. The views and opinions expressed herein do not necessarily reflect those of the European Commission or of the ITER Organization. This work was partially funded by the Academy of Finland project No. 324759. We acknowledge the computational resources provided by Aalto Science-IT project.
\\
\\
This is the Accepted Manuscript version of an article accepted for publication in Nuclear Fusion. IOP Publishing Ltd is not responsible for any errors or omissions in this version of the manuscript or any version derived from it. The Version of Record is available online at https://doi.org/10.1088/1741-4326/abdfdd.

\section*{References}
\bibliographystyle{unsrt}
\bibliography{collection}

\end{document}